# Automatic detection of abnormal clinical EEG: comparison of a finetuned foundation model with two deep learning models

Aurore Bussalb[1 *], François Le Gac[1], Guillaume Jubien[1], Mohamed Rahmouni[1], Ruggero G. Bettinardi[1], Pedro Marinho R. de Oliveira[1], Phillipe Derambure[2, 3], Nicolas Gaspard[4, 5, 6], Jacques Jonas[7, 8], Louis Maillard[7, 8], Laurent Vercueil[9], Hervé Vespignani[1], Philippe Laval[10], Laurent Koessler[1, 7], Ulysse Gimenez[1]

[1]BioSerenity, Paris, France
[2]Université de Lille, Lille Neuroscience & Cognition | UMRS1172
[3]Neurophysiologie Clinique, CHU de Lille
[4]Service de Neurologie, Hôpital Universitaire de Bruxelles - Bruxelles, Belgique
[5]Laboratoire de Neurologie Expérimentale, Université Libre de Bruxelles - Bruxelles, Belgique
[6]Department of Neurology, Yale University School of Medicine - New Haven, CT USA
[7]Université de Lorraine, IMoPA, UMR CNRS 7365, Nancy, France
[8]Université de Lorraine, CHRU-Nancy, Service de Neurologie, F-54000 Nancy, France
[9]EFSN CHU Grenoble Alpes/LPNC Université Grenoble Alpes, France
[10]Jolt Capital, Paris, France

*Corresponding author (aurore.bussalb@bioserenity.com)

## Abstract

Electroencephalography (EEG) is commonly used by physicians for the diagnosis of numerous neurological disorders. Due to the large volume of EEGs requiring interpretation and the specific expertise involved, artificial intelligence-based tools are being developed to assist in their visual analysis. In this paper, we compare two deep learning models (CNN-LSTM and Transformer-based) with BioSerenity-E1, a recently proposed foundation model, in the task of classifying entire EEG recordings as normal or abnormal. The three models were trained or finetuned on 2,500 EEG recordings and their performances were evaluated on two private and one public datasets: a large multicenter dataset annotated by a single specialist (dataset A composed of n = 4,480 recordings), a small multicenter dataset annotated by three specialists (dataset B, n = 198), and the Temple University Abnormal (TUAB) EEG corpus evaluation dataset (n = 276). On dataset A, the three models achieved at least 86% balanced accuracy, with BioSerenity-E1 finetuned achieving the highest balanced accuracy (89.19% [88.36-90.41]). BioSerenity-E1 finetuned also achieved the best performance on dataset B, with 94.63% [92.32-98.12] balanced accuracy. The models were then validated on TUAB evaluation dataset, whose corresponding training set was not used during training, where they achieved at least 76% accuracy. Specifically, BioSerenity-E1 finetuned outperformed the other two models, reaching an accuracy of 82.25% [78.27-87.48]. Our results highlight the usefulness of leveraging pre-trained models for automatic EEG classification: enabling robust and efficient interpretation of EEG data with fewer resources and broader applicability.

# 1. Introduction

Electroencephalography (EEG) is a widely used tool to help for the diagnosis of numerous neurological disorders (Sanei et al., 2013, André-Obadia et al., 2014), such as epilepsy (Tatum et al., 2018), and brain tumors (Bera, 2021). Because of its high temporal resolution, non-invasiveness, portability, and relatively low cost, the volume of EEG is immense, but it requires specific expertise to interpret it (Kwon et al. 2022, Nascimento & Gavvala, 2021), making it challenging to review visually all the recordings (Smith et al., 2005, Subasi, 2020). In addition, low signal to noise ratio (due to high skull resistivities) and mixed electrical brain sources (volume conduction) can blur the scalp EEG biomarkers to detect (Koessler et al., 2015). Thus, developing artificial intelligence (AI) based methods to help physicians in their analysis is an important and useful research field.

AI has already been applied to a large spectrum of neurological disorders: Alzheimer's disease detection (Puri et al., 2023), sleep disorder detections (Toma & Choi, 2023), epileptic seizure prediction, detection, and classification (Farooq et al., 2023, Zambrana-Vinaroz et al. 2022, Albaqami et al., 2023b, Jing et al., 2023, Jing et al., 2020, Barnett et al., 2024), and detecting if an EEG is abnormal (Tveit et al., 2023, S. Roy et al., 2018, Albaqami et al., 2023a). An EEG is considered abnormal if it has findings known to be associated with a pathologic or disease state (Libenson, 2024, Britton et al., 2016). Examples of findings include generalized delta slowing which can be associated with many abnormal states (e.g., coma, post-seizure state, meningitis, anesthesia) or epileptiform discharges.

To automatically analyze EEGs, different approaches have been implemented over the years: extracting features from the signals then feeding them to a machine learning algorithm, or using an end-to-end approach with deep learning (Craik et al., 2019, Gemein et al., 2020). In the case of automatic abnormal EEG classification, both strategies have been investigated (Albaqami et al., 2021, Tveit et al., 2023), but in recent years researchers have tended to focus on the deep learning approach (Craik et al., 2019, Y. Roy et al. 2019). Several model architectures have already been tested with promising results (Albaqami et al., 2023a, Tveit et al., 2023, Alhussein et al., 2019, S. Roy et al. 2018).

In this study, our main objective was to compare a finetuned foundation model with two separate deep learning models trained from scratch on the task of classifying 20 minutes EEG recordings as normal or abnormal. Specifically, a new foundation model, BioSerenity-E1 (Bettinardi et al., 2025), pretrained using masked-self-supervised learning on clinical EEG recordings, was finetuned on 2,500 EEGs. It was compared with two deep learning models (a convolutional neural network (CNN) coupled with a Long Short-Term Memory (LSTM) and a Transformer-based model), which were implemented and trained from scratch on the same 2,500 EEG recordings.

We tested our models on three distinct datasets to demonstrate their robustness and generalizability. The first dataset is a large multicenter dataset (n = 4,480 recordings) annotated by one expert, which capture a wide variability of participants indications and recording equipment. Then, a second small dataset (n = 198) annotated by three experts was used. Finally, we benchmarked our algorithms using the public dataset Temple University Hospital Abnormal (TUAB) EEG Corpus evaluation dataset (n = 276) (Lopez de Diego, 2017).

Furthermore, the impact of the training size on the models' performances was investigated: the three models were also trained or finetuned using progressively smaller datasets (from 1250 to 74 recordings).

## 2. Materials and methods

### a. Description of the datasets

The EEGs included in this research were acquired in routine or intensive care unit settings, for various indications such as epilepsy, dementia, coma, or stroke. These recordings were 20 minutes EEGs collected from adult participants (≥18 years old) with the 19 scalp electrodes placed according to the 10/20 system (Klem, 1999). EEGs with some non-functional or artifactual channels were discarded. Data were anonymized and participants agreed to the use of their personal data for research and development purposes.

All the datasets used are listed in Table 1 and further described in the following subsections.

| Dataset name | Scope | Number of recordings | Percentage of abnormal recordings | Duration (minutes) | Age (years) |
|---|---|---|---|---|---|
| Training dataset | Training/fine-tuning | 2,500 | 50 | 20.5 ± 1.5 | 60.0 ± 21.0 |
| Dataset A | Test | 4,480 | 42.21 | 20.4 ± 1.1 | 58.9 ± 22.0 |
| Dataset B | Test | 198 | 76.26 | 20.4 ± 1.1 | 63.3 ± 22.1 |
| TUAB evaluation dataset | Test | 276 | 54.35 | 22.4 ± 5.2 | 51.0 ± 18.4 |

*Table 1:* Summary of all datasets used in this research. Models are trained/finetuned on Training dataset. Models were all tested on the three datasets.

### i. Training dataset

This dataset consisted of 2,500 EEGs from Neurophy, the tele-interpretation service of BioSerenity (Paris, France), that were collected with three different EEG equipment namely Micromed®, Neuronaute®, and Nihon Kohden, in more than one thousand hospitals and clinics in France between 2020 and 2023. The mean EEG duration of this dataset was 20.5 ± 1.5 minutes and the sampling frequency was either 250, 256, or 500 Hz. The mean age was 60.0 ± 21.0 years.

EEG recordings were visually reviewed by one physician in current practice who provided a written conclusion on each recording The dataset was perfectly balanced between normal and abnormal recordings.

ii. Test datasets

**A large multicenter dataset: dataset A**

The dataset consisted of 4,480 EEGs recordings collected from about 500 centers in France between 2020 and 2023 through Neurophy, ensuring diversity in acquisition settings and equipment as multiple acquisition systems from different vendors (Micromed, Bioserenity, and Nihon Kohden) were used. Among these, 2,589 (57.79%) were labelled as normal by a neurologist. The annotation process was identical to that used for the training dataset. The mean age of participants was 58.9 ± 22.0 years. The mean EEG duration was 20.4 ± 1.1 minutes and the sampling frequency was either 250, 256, or 500 Hz.

**A small multicenter and rigorously validated dataset: dataset B**

EEGs were collected from about 60 different hospitals in France through Neurophy, between October and November 2024. This dataset only included EEGs recorded with Neuronaute® equipment. Recordings where participants were under sedation were discarded.

Each recording was reviewed by three different qualified physicians in two steps. The first review was made by a physician in current practice, who was not the same for all the recordings. Secondly, three groups of two different physicians were arbitrary defined: each group was assigned to one third of the total number of recordings and reviewed them blindly, anonymously, and independently. The consensus of the three scorings was then computed: it was obtained by the 2/3 majority, recordings for which no consensus was reached were excluded.

At the end of the review process, this dataset consisted of 198 recordings: the mean age of subjects was 63.3 ± 22.1 years, the mean duration of the EEG recordings was 20.4 ± 1.1 minutes, and 76.26% records were abnormal. The sampling frequency of all the recordings was 250 Hz.

**A publicly available dataset: TUAB evaluation dataset**

Most studies that aimed at identifying if an EEG was normal or abnormal relied on TUAB (Lopez de Diego, 2017, Obeid & Picone., 2016, Albaqami et al., 2023a, Alhussein et al., 2019, S. Roy et al., 2018), a publicly available database. TUAB v2.0.0 was divided into a training and an evaluation datasets: in our analysis we only used the evaluation set to be able to compare our results to the literature and assess if our algorithms were robust and able to generalize. The labeling of recordings was caried out by neurologists and students.

The evaluation set was composed of 276 adults EEG recordings: 150 were abnormal (54.35%). The mean EEG duration was 22.4 ± 5.2 minutes and the sampling frequency was either 250 or 256 Hz. The mean age was 51.0 ± 18.4 years.

b. Train two deep learning models from scratch

i. Preprocessing

We applied a fifth-order Butterworth highpass filter with a 0.5 Hz cutoff frequency on the 19 channels EEG of the 10/20 system to reduce low-frequency drift, followed by an IIR notch filter at 50 Hz (or 60 Hz for recordings from TUAB evaluation dataset) to remove power-line interference. Since some datasets included EEGs recorded with different equipment, acquisition references were not always the same so an average re-referencing was performed. Lastly, all signals were resampled at 256 Hz.

ii. The CNN-LSTM model

CNN and LSTM have often been used in EEG classification: they have been applied separately, but also in a hybrid way (Craik et al., 2019). Based on the promising results achieved by similar architectures in normal/abnormal classifications (Khan et al., 2022, Albaqami et al., 2023a) and in other EEG areas such as in human emotion recognition (Li et al., 2017), we choose to implement a model with the hybrid approach.

While in several previous studies just a part of the signal was given (Albaqami et al., 2023a, S. Roy et al., 2018, Schirrmeister et al., 2017), we decided to provide the whole EEG as input of the model. This choice was motivated by the fact that we had one label per recording and we had no information regarding the presence of the abnormalities on the entire signal. Additionally, although many neural networks, especially CNNs, take EEG spectrograms or Fourier transforms as input (Alhussein et al., 2019, Vrbancic & Podgorelec, 2018, Kuanar et al., 2018), we choose here to give the EEG preprocessed as described in the previous section as input as it was done in Albaqami et al. (2023a).

The model was developed using Tensorflow version 2.16.1 and Python 3.11: it was composed of a succession of four one-dimensional convolutional layers separated by average pooling layers. A LSTM layer and a fully connected layer followed. We applied dropout with a probability of 0.5 right after the LSTM layer. Figure 1 presents the architecture of the model.

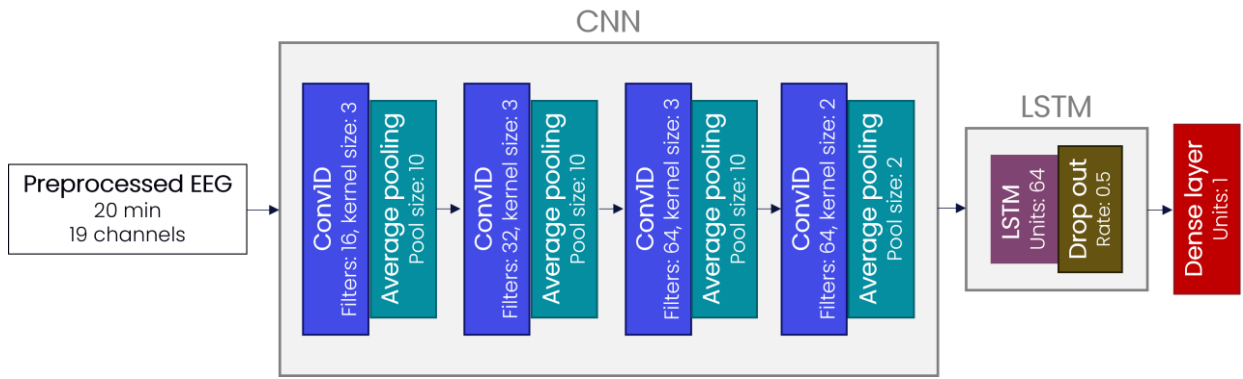

*Figure 1:* Architecture of the CNN-LSTM model.

The convolutional layers aimed at extracting features from the EEG: we made the number of filters increase with the depth of the model, to ensure to extract more subtle features in the deeper layers. LSTM layers are known for their ability to handle time-dependent information and temporal patterns and are well suited for EEG (Li et al., 2022, Song et al., 2022).

The F1-score was computed on the validation dataset (corresponding to 20% of the training set) for each training epoch and was the value to monitor during the training step. Early stopping with a patience of 5 epochs was applied.

The model returned the probability for a recording to be abnormal: a threshold set to 0.5 enabled to classify the record as abnormal (probability > 0.5) or normal.

iii. The Transformer model

The second model consisted of a Short-Time Fourier Transform (STFT) followed by a 2D-CNN and a transformer block. Given the multidimensional nature and potential long size of EEGs, we first applied STFT to compress the raw signal. STFT segments the EEG signal into small windows and applies the Fourier transform to each, converting the time-domain signal into the time-frequency domain. This method, widely used in signal processing (Ma et al., 2023, Peng et al., 2022), provides a compact representation while preserving spectral and temporal information. We applied STFT independently to each EEG channel using windows of 1s seconds with 10% overlapping.

The resulting spectrograms were processed by a CNN which performed 2D convolutions over the time axis. The CNN extracts local features while reducing time resolution, producing embeddings that further compress the signal representation. Varying the temporal kernel sizes allows CNNs to capture frequency components at different scales (Lawhern et al., 2018). Concatenating the embeddings across channels preserves spatial relationships in the data.

While CNNs effectively capture local patterns, they do not explicitly model global dependencies. To address this, a transformer block followed the 2D-CNN. Transformers are widely used in sequential data processing (Dong et al., 2018, Wan et al., 2023) as they efficiently capture long-range dependencies using the attention mechanism. They are also highly effective as they process all time steps in parallel. A classifier head finally used the transformer block output to classify the EEG as normal or abnormal. The architecture of the model is presented in Figure 2.

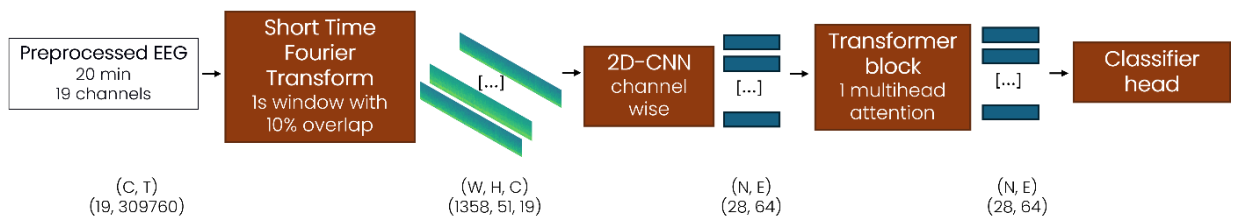

*Figure 2:* Transformer model architecture, where C corresponds to the number of channels, T to the number of time points, W to the width of the spectrogram, H to the height of the spectrogram, N to the sequence length, E to the embedding dimension.

### c. Finetune a foundation model

#### i. Preprocessing

The preprocessing of the EEGs for this second strategy was slightly different from the preprocessing of the two previous algorithms to comply with the preprocessing used during the self-supervised training of BioSerenity-E1. EEG signals were first filtered: a fifth-order Butterworth highpass filter with a 0.5 Hz cutoff frequency was applied, followed by a fifth-order Butterworth lowpass filter with a 45 Hz cutoff frequency, and an IIR notch filter at 50 Hz (or 60 Hz for recordings from TUAB evaluation dataset). Signals were then resampled at 128 Hz, and 16 channels were kept (Fz, Cz, and Pz were discarded) and arranged in a specific order to preserve spatial information as required by the pretrained model. Lastly, signals were average re-referenced.

#### ii. Description of BioSerenity-E1 finetuned

A foundation model is a large pretrained model using self-supervised learning that serves as a base for various downstream tasks. It is trained on vast amounts of data and can be finetuned for specific applications (Bommasani et al., 2021). In our study, we finetuned BioSerenity-E1, a foundation model already pretrained on 4,000 hours of EEG in a self-supervised fashion using masked token prediction to learn how EEGs were structured by reconstructing them. In the rest of the article, we will refer to this finetuned version of BioSerenity-E1 as "BioSerenity-E1 finetuned".

BioSerenity-E1 have already been successfully used to classify 16-second windows of EEG into normal or abnormal, showing high performances (Bettinardi et al. 2025). In the current work, we focused on adapting BioSerenity-E1 to get predictions at the level of the whole EEG recordings. To do so, each recording was first split into non-overlapping 16-second windows, because BioSerenity-E1 took inputs of 16 seconds by 16 channels. The resulting vector shape was ($N_W$, $N_S$, $N_C$), with $N_W$ the number of 16-second windows in the recording, $N_S$ the number of samples per window, and $N_C$ the number of channels. This vector was the input of the pretrained model, whose weights were frozen, and features for each window was computed: we obtained a vector whose shape was ($N_W$, $N_P$, $D_E$), with $N_P$ the number of vectors storing the representation of the EEG signal corresponding to one second and one channel, and $D_E$ the embedding dimension. These features were then averaged per recording to get a ($N_P$, $D_E$) vector, which was fed to a prediction head. The prediction head was a succession of three convolutional layers separated by batch normalization and dropout with a probability of 0.4. A global average pooling layer was then added, followed by several linear transformations separated by dropout with a probability of 0.4. It was finetuned using Pytorch 2.4 and Python 3.10 (Figure 3).

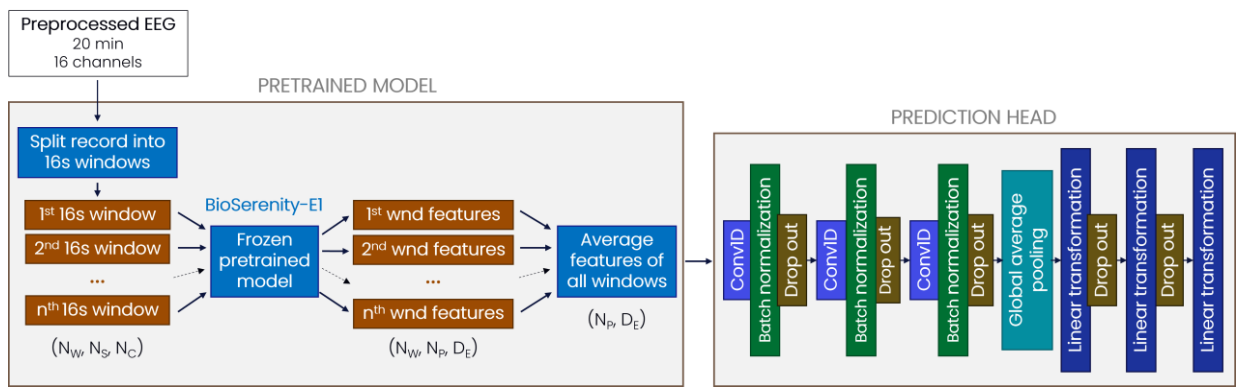

*Figure 3:* Architecture of BioSerenity-E1 finetuned. The frozen pretrained model (BioSerenity-E1) extracted features that were given to the trainable prediction head. $N_W$ corresponds to the number of windows, $N_S$ to the number of samples inside a window, $N_C$ to the number of channels, $N_P$ to the number of vectors storing the representation of the EEG signal corresponding to one second and one channel, and $D_E$ to the embedding dimension.

The F1-score was computed on the validation dataset (corresponding to 20% of the training set) for each training epoch and was the value to monitor. Early stopping with a patience of 5 epochs was applied.

BioSerenity-E1 finetuned returned the probability for a recording to be abnormal: a threshold set to 0.5 enabled to classify the record as abnormal (probability > 0.5) or normal.

### d. Outcome measures

#### i. Analysis of the consensus

To measure the reliability of the physicians' labeling of dataset B during the second step of the review process, the Cohen's kappa was used (McHugh, 2012, Cohen, 1960). This metric measures the strength of agreement between two scorers beyond chance. In our analysis, it was computed to assess:

- The agreement between each of the six physicians and the consensus,
- The agreement between each algorithm's predictions and the consensus.

The Cohen's kappa was interpreted according to Landis and Koch criteria (Landis & Koch, 1977). Its 95% confidence interval was computed by bootstrap resampling (n = 1,000 resamples). The comparisons of kappa scores were based on their 95% confidence interval: the difference between two kappa scores was considered statistically significant if there was no overlap.

  ii. Performance of the algorithms

The performance of the algorithms was assessed by three metrics on the three test datasets described previously (Parikh et al., 2008):

- Sensitivity (also called true positive rate), which corresponds to the ability of the algorithm to correctly classify a record as abnormal,
- Specificity, which corresponds to the ability of the algorithm to correctly classify a record as normal,
- Balanced accuracy, which corresponds to the arithmetic mean of sensitivity and specificity, and is robust to class imbalance. To compare precisely our models to previous published results, the accuracy was provided instead for TUAB evaluation dataset.

For each of these metrics, its 95% confidence interval was computed by bootstrap resampling (n = 1,000 resamples).

For dataset B, a Receiver Operating Characteristic (ROC) curve was plotted for each algorithm, and their Area Under the Curve (AUC) was given as well.

Balanced accuracies (or accuracies for TUAB evaluation dataset) of the three models on each test dataset were bootstrap resampled (n=1,000 resamples) and were then statistically compared thanks to an ANOVA (Sthle & Wold, 1989). Statistical significance was set at p-value < 0.05.

  e. Analysis of the impact of training size

The impact of the training size was also explored by analyzing the three models performances after decreasing their training dataset size gradually. To do so, five new training datasets, described in Table 2, were created from Training dataset: recordings were randomly selected from this dataset, the only constraint was the datasets had to be balanced.

| Dataset name | Number of recordings | Percentage of abnormal recordings | Duration (minutes) | Age (years) |
|---|---|---|---|---|
| Subset 1250 | 1250 | 50 | 20.5 ± 1.6 | 60.2 ± 20.9 |
| Subset 750 | 750 | 50 | 20.4 ± 1.2 | 59.7 ± 21.0 |
| Subset 250 | 250 | 50 | 20.5 ± 1.2 | 57.1 ± 22.0 |
| Subset 126 | 126 | 50 | 20.5 ± 1.3 | 60.8 ± 19.9 |
| Subset 74 | 74 | 50 | 20.4 ± 1.5 | 58.4 ± 21.2 |

*Table 2:* Description of datasets created to investigate the impact of decreasing the training size on models' performance.

To get more reliable results, each algorithm was trained on each dataset and on Training dataset three times. Models were then tested on dataset B: balanced accuracies of same models trained on the same dataset were averaged and their standard deviation was computed.

## 3.  Results

### a.  Assessment of the consensus

Cohen's kappa between each of the six physicians from the second annotation phase and the consensus with their 95% confidence interval are given in Table 3.

| Physician | Cohen's kappa (95% CI) |
|---|---|
| A | 0.72 [0.56-0.87] |
| B | 0.71 [0.53-0.86] |
| C | 0.74 [0.59-0.86] |
| D | 0.88 [0.74-0.97] |
| E | 0.70 [0.56-0.84] |
| F | 0.81 [0.68-0.92] |

*Table 3:* Cohen's kappa with their 95% confidence interval between physicians from the second phase of the annotation process and the consensus.

Cohen's kappa were between 0.61 and 0.80 for 4 physicians, indicating a substantial agreement. The agreement of two physicians with the consensus was almost perfect (i.e., Cohen's kappa ≥ 0.81). There was no significant statistical difference between the kappa scores. The average kappa score was 0.76 with a 95% confidence interval of [0.70-0.88], which was computed by bootstrap resampling (n = 1,000 resamples).

### b.  Validation on dataset A

Sensitivity, specificity, and balanced accuracy with their respective 95% confidence interval were computed for each proposed algorithm on dataset A and the results are summarized in Figure 4.

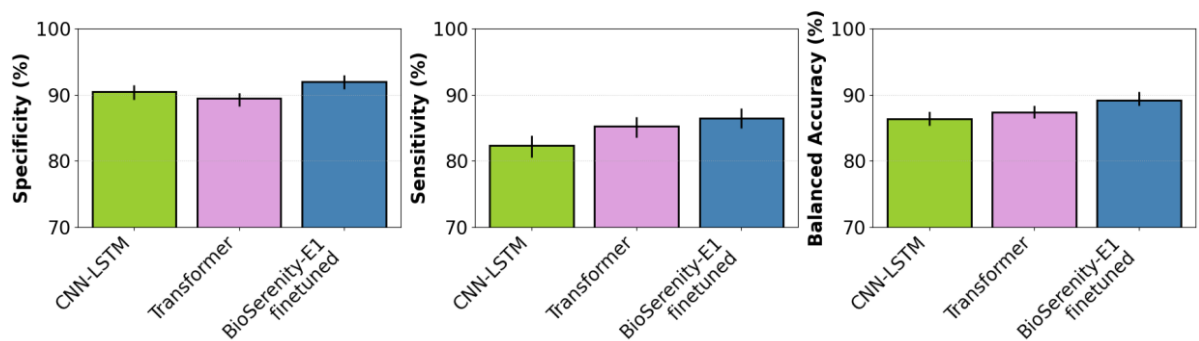

*Figure 4:* Specificity, sensitivity, and balanced accuracy with their respective 95% confidence interval on dataset A for each proposed algorithm.

Performances on this dataset were high for the three algorithms: the CNN-LSTM model presented the lowest balanced accuracy (86.34% [85.36-87.42]). BioSerenity-E1 finetuned reached the best specificity, sensitivity, and balanced accuracy (respectively 91.93% [90.83-93.00], 86.46% [84.94-87.99], and 89.19% [88.36-90.41]).

There was a significant statistical difference between the balanced accuracies of the three models (p-value around 10e-5).

### c. Validation on dataset B

Sensitivity, specificity, balanced accuracy, and Cohen's kappa with the consensus with their respective 95% confidence interval were computed for each proposed algorithm on dataset B. The results are summarized in Table 4.

|  | **Specificity (%)** | **Sensitivity (%)** | **Balanced accuracy (%)** | **Cohen's kappa** |
|---|---|---|---|---|
| CNN-LSTM model | 95.74 [88.64-100] | 80.79 [74.51-96.79] | 88.27 [84.10-92.33] | 0.64 [0.52-0.75] |
| Transformer model | 95.74 [88.64-100] | 86.09 [80.39-93.47] | 90.92 [90.34-91.14] | 0.72 [0.61-0.82] |
| BioSerenity-E1 finetuned | **97.87** [93.33-100] | **91.39** [86.71-95.65] | **94.63** [92.32-98.12] | **0.82** [0.73-0.91] |

*Table 4:* Specificity, sensitivity, balanced accuracy, and Cohen's kappa with the consensus with their respective 95% confidence interval on dataset B for each proposed algorithm. The highest performances are in bold.

All of the three algorithms showed high performances, but their balanced accuracies are statistically different (p-value around 10e-5). The highest results were achieved by finetuning BioSerenity-E1 (94.63% [92.32-98.12] balanced accuracy), and its agreement with the consensus was considered almost perfect (0.82). There was no statistical difference between the Cohen's kappa between each physician and the consensus, listed in Table 3, and the Cohen's kappa between each algorithm's predictions and the consensus based on their 95% confidence interval. Besides, only physician B presented a higher Cohen's kappa than BioSerenity-E1 finetuned.

The ROC curves of the models along with their AUC highlighted how close the performances of the algorithms were, but with a better performance for BioSerenity-E1 finetuned (AUC = 0.98, Figure 5).

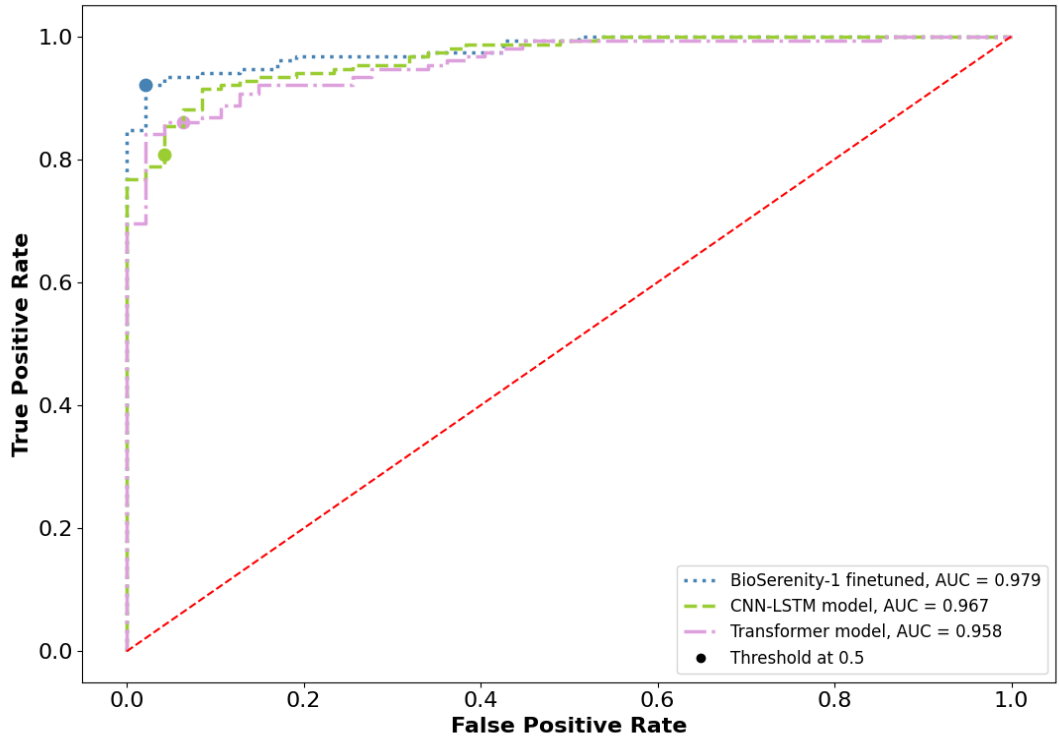

*Figure 5:* ROC curves of the three models with their AUC on dataset B.

### d. Validation on TUAB evaluation dataset

The leaderboard of the recent previous studies using the publicly available TUAB dataset to develop their deep learning models is summarized in Table 5 and in Figure 6. The models from the reported studies were all trained and tested on the corresponding train and test sets of TUAB, whereas the CNN-LSTM model, Transformer model, and BioSerenity-E1 finetuned were trained on an independent set of data (see "Training Dataset" subsection) and only tested using TUAB evaluation dataset.

|  | **Trained on TUAB training set** | **Specificity (%)** | **Sensitivity (%)** | **Accuracy (%)** |
|---|---|---|---|---|
| S. Roy et al. (2018) | ✅ | N/A | N/A | 82.27 [N/A] |
| S. Roy et al. (2019) | ✅ | N/A | N/A | 86.6 [N/A] |
| Amin et al. (2019) | ✅ | 94.67 [N/A] | 78.57 [N/A] | 87.32 [N/A] |
| Alhussein et al. (2019) | ✅ | **96.57** [N/A] | 80.16 [N/A] | **89.13** [N/A] |
| Yildirim et al. (2020) | ✅ | N/A | N/A | 79.34 [N/A] |
| Gemein et al. (2020) | ✅ | 91.60 [N/A] | 79.70 [N/A] | 86.16 [N/A] |
| Khan et al. (2022) | ✅ | N/A | N/A | 85 [N/A] |
| Kiessner et al. (2023) | ✅ | 91.53 [N/A] | 78.81 [N/A] | 85.72 [N/A] |
| Albaqami et al. (2023a) | ✅ | 92.00 [N/A] | **84.92** [N/A] | 88.76 [N/A] |
| CNN-LSTM model | ❌ | 94.66 [90.79-97.87] | 53.97 [45.08-62.50] | 76.09 [71.26-81.36] |
| Transformer model | ❌ | 91.33 [86.89-95.49] | 62.69 [54.69-71.65] | 78.26 [73.55-82.97] |
| BioSerenity-E1 finetuned | ❌ | 96.00 [92.62-98.74] | 65.87 [56.91-73.28] | 82.25 [78.27-87.48] |

*Table 5:* Comparison of the proposed models trained and other state-of-the-art methods on TUAB evaluation dataset: their specificity, sensitivity, and accuracy with their respective 95% confidence interval are provided when available. The highest performances are in bold. All benchmark models were trained using TUAB training set, whereas CNN-LSTM, the Transformer model, and BioSerenity-E1 finetuned were trained/finetuned on a different set of EEG records, and externally validated on TUAB evaluation dataset.

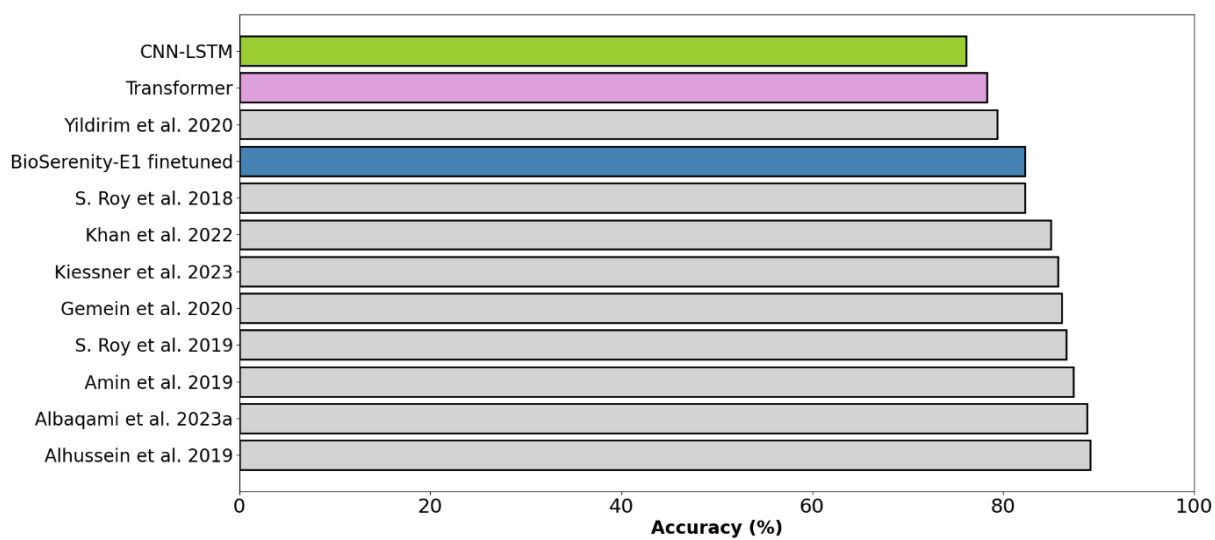

*Figure 6:* Accuracies of the proposed models (pink for Transformer model, green for CNN-LSTM, and blue for BioSerenity-E1 finetuned) and other state-of-the-art methods on TUAB evaluation dataset. All benchmark models were trained using TUAB training set, whereas CNN-LSTM, Transformer model, and BioSerenity-E1 finetuned were trained/finetuned on a different set of EEG recordings, and externally validated on TUAB evaluation dataset.

Our algorithms behaved in a similar way on TUAB evaluation dataset: high specificities, almost as high as the one from Alhussein et al., 2019 for BioSerenity-1 finetuned (96.00% [92.62-98.74]); low sensitivities, lower than the ones given in the previous published articles; accuracies lower from the literature as shown in Figure 6, except for BioSerenity-E1 finetuned (82.25% [78.27-87.48]).

There was a significant statistical difference between the balanced accuracies of the three models (p-value around 10e-5).

### e. Impact of the training dataset size

Average balanced accuracy for each model trained on a specific number of records were computed on dataset B and are presented in Figure 7.

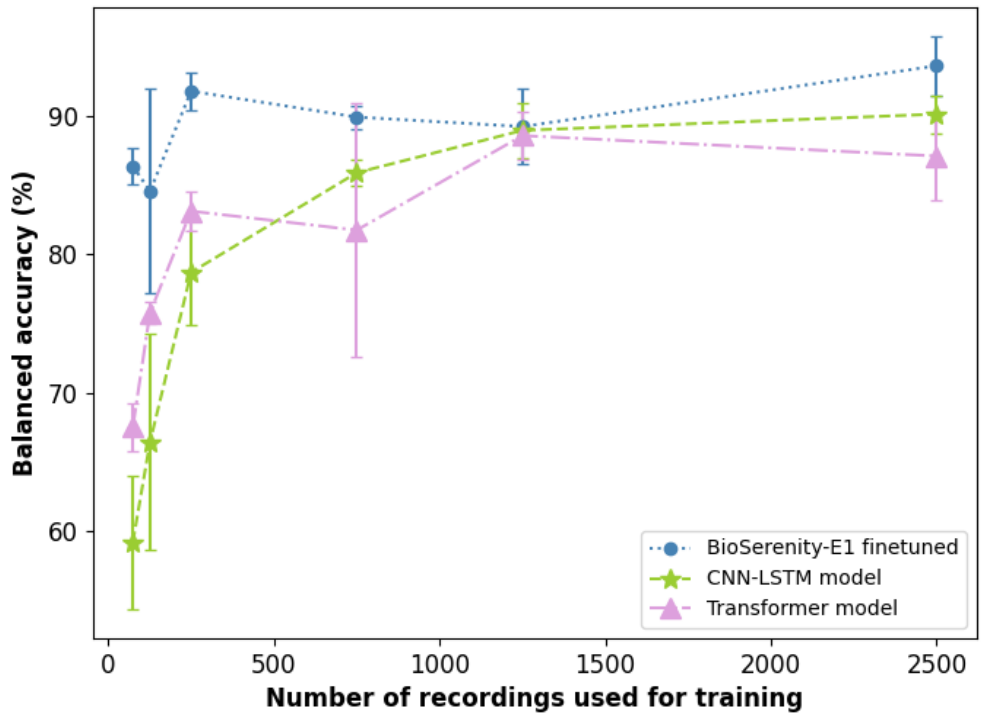

*Figure 7:* Average balanced accuracies and their standard deviation on dataset B of three trainings for each model according to training size.

Performances of BioSerenity-E1 finetuned were always higher than CNN-LSTM and Transformer models regardless of the training size. For training sizes ≥ 1250 recordings (corresponding to > 416 total hours of EEG), all the three models achieved satisfactory performances, but for lower training sizes (n ≤ 250 recordings, corresponding to < 85 EEG hours) the average balanced accuracies of the CNN-LSTM and Transformer models were far lower than the ones of BioSerenity-E1 finetuned, which were still high (86.36% ± 1.27 for n = 74, 84.56% ± 7.36 for n = 126, and 91.78% ± 1.38 for n = 250). Notably, BioSerenity-E1 finetuned resulted in a 27.94% increase in balanced accuracy compared to the Transformer model using as few as 74 recordings (corresponding to approximately 24 hours of EEG data).

## 4. Discussion

In this research, we compared two conventional algorithms and BioSerenity-E1 finetuned, which aimed at classifying EEG recordings into normal or abnormal. The three models were trained/finetuned on a multicenter and balanced dataset containing routine and intensive care unit EEGs recorded with different equipment. Models' performance was internally validated using two separate sets of records from proprietary datasets (datasets A and B) and externally validated on the test set of a well-known public database (TUAB evaluation dataset). Results demonstrated the algorithms' robustness and ability to generalize. The three models were thoroughly compared between each other, but also to the literature through the TUAB evaluation dataset. The most notable innovation presented in this study lies in the strategic application of a foundation model for automated EEG classification.

The diagnostic accuracy of these models showed high performances across our 3 test datasets, namely a large private multicenter with different equipment dataset (dataset A, n = 4,480), a small private multicenter dataset labeled by 3 different experts (dataset B, n = 198), and the TUAB evaluation dataset (n = 276). Regarding dataset B, the agreement between each physician and the consensus was at least substantial and almost perfect for two physicians, making the consensus reliable. The results on TUAB evaluation dataset indicated that the models detected correctly normal records, but had more trouble when it came to abnormal records: this finding may be explained by the fact that these kinds of abnormalities were not in our training set.

BioSerenity-E1 finetuned outperformed the CNN-LSTM model and Transformer model on the three datasets showing that the self-supervised pretraining of BioSerenity-E1 enabled it to effectively capture the intrinsic structure of EEG signals without relying on manually labeled data.

To automatically classify EEG recordings into normal and abnormal, several deep learning models were already developed and described in the literature (S. Roy et al., 2018, S. Roy et al., 2019, Amin et al., 2019, Alhussein et al., 2019, Yildirim et al., 2020, Gemein et al., 2020, Khan et al., 2022, Kiessner et al., 2023, Albaqami et al., 2023a). Most of them took as input the EEG signal, but sometimes EEG features could be given such as Fast Fourier Transform computed in specific bands of the signal (Alhussein et al., 2019). In our case, BioSerenity-E1 finetuned and CNN-LSTM model took the temporal EEG as input while the Transformer model took the Short Time Fourier Transform.

The models' architecture were variable, but CNN were often used (S. Roy et al., 2018, S. Roy et al., 2019, Amin et al., 2019, Yildirim et al., 2020, Khan et al., 2022, Kiessner et al., 2023), and were in some cases coupled with a LSTM (Khan et al., 2022), that's why the three models presented in this article had at least CNN layers.

Models described in the literature were mainly only trained on TUAB training set, but some used other datasets as well (Alhussein et al., 2019). Their performances were often only available for the corresponding test dataset (TUAB evaluation dataset), preventing a conclusion about their generalizability. That's why, to strongly validate their results, Albaqami et al. (2023a) gave performances on an entirely separate dataset, Temple University Hospital EEG Epilepsy Corpus (Veloso et al., 2017), without further hyperparameter tuning or adjustments. We followed the same strategy by evaluating our models on TUAB evaluation dataset.

The balanced accuracy scores of BioSerenity-E1 finetuned on datasets A and B were superior to the scores reported in Table 5. Additionally, they were close to the ones claimed by SCORE-AI algorithm (Tveit et al., 2024), which achieved 95.00% (95% confidence interval, [89.61-97.88]) accuracy on 100 EEGs evaluated by 11 experts while predicting if an EEG was normal or not. To the best of our knowledge, only SCORE-AI algorithm had shown clinical utility for the diagnosis of normal vs abnormal EEG recordings, and had been integrated with the Natus® NeuroWorks® EEG reader. This model presented a CNN architecture that classified the EEG into 5 classes (normal, epileptiform-focal, epileptiform-generalized, nonepileptiform-diffuse, and nonepileptiform-focal), and was trained on a multicenter dataset of 30,493 records.

Our algorithms were neither trained on neonates nor children and therefore should not be used for discriminating normal vs abnormal recordings in this population. Another limit was the restricted degree of interpretation of their output: although our methods provided a diagnostic, they neither gave any information regarding the type of abnormality nor specifically indicated what lead to the diagnostic. For the later point, several methods were published in order to provide more information on what triggered the neural network decision such as heatmaps (Storås al., 20223) or Saliency-Based Methods (Gomez et al., 2022), which may help physicians to understand the output of the model and make it more trustworthy (Rajpurkar et al., 2022). Finally, specific details about the abnormalities present in the recordings were not available in our datasets. This shortcoming impacted the analysis of our current results: knowing the distribution of abnormalities inside the different datasets may had help to figure out why the algorithms failed to predict the correct class for some recordings.

## Conclusion

In this research, we proposed a new model, BioSerenity-E1 finetuned, to classify EEGs into normal or abnormal that reached high performances. By finetuning BioSerenity-E1 on a relatively modest dataset of 2,500 EEG recordings, we demonstrated higher performance compared to conventional deep learning approaches trained entirely from scratch on the same dataset. Additionally, despite variations in EEG acquisition conditions, equipment, and recording environments, BioSerenity-E1 finetuned demonstrated excellent robustness, indicating its potential broad applicability across diverse clinical settings.

Through our study, we highlighted that leveraging pretrained foundation models substantially decreased the required amount of labeled data, facilitated faster training times, reduced computational costs, and lead to greater accessibility to high-performance EEG classification tools. In the future, we will focus on further developing BioSerenity-E1 and assess its performance on other relevant downstream tasks and datasets.

Bussalb et al., 2025Bettinardi, R. G., Rahmouni, M., Gimenez, U. (2025). BioSerenity-E1: a self-supervised EEG model for medical applications. *arXiv preprint arXiv:2503.10362*.

Jing, J., Ge, W., Hong, S., Fernandes, M. B., Lin, Z., Yang, C., ... & Westover, M. B. (2023). Development of expert-level classification of seizures and rhythmic and periodic patterns during EEG interpretation. *Neurology*, *100*(17), e1750-e1762.

Jing, J., Sun, H., Kim, J. A., Herlopian, A., Karakis, I., Ng, M., ... & Westover, M. B. (2020). Development of expert-level automated detection of epileptiform discharges during electroencephalogram interpretation. *JAMA neurology*, *77*(1), 103-108.

Barnett, A. J., Guo, Z., Jing, J., Ge, W., Kaplan, P. W., Kong, W. Y., ... & Westover, M. B. (2024). Improving clinician performance in classifying EEG patterns on the ictal–interictal injury continuum using interpretable machine learning. *NEJM AI, 1*(6), AIoa2300331.

André-Obadia, N., Sauleau, P., Cheliout-Heraut, F., Convers, P., Debs, R., Eisermann, M., ... & Lamblin, M. D. (2014). Recommandations françaises sur l'électroencéphalogramme. *Neurophysiologie Clinique/Clinical Neurophysiology*, *44*(6), 515-612.

Bera, T. K. (2021, March). A review on the medical applications of electroencephalography (EEG). In *2021 Seventh International conference on Bio Signals, Images, and Instrumentation (ICBSII)* (pp. 1-6). IEEE.

Koessler, L., Cecchin, T., Colnat-Coulbois, S., Vignal, J. P., Jonas, J., Vespignani, H., ... & Maillard, L. G. (2015). Catching the invisible: mesial temporal source contribution to simultaneous EEG and SEEG recordings. *Brain topography*, *28*, 5-20.

Britton, J. W., Frey, L. C., Hopp, J. L., Korb, P., Koubeissi, M. Z., Lievens, W. E., ... & St Louis, E. K. (2016). Electroencephalography (EEG): An introductory text and atlas of normal and abnormal findings in adults, children, and infants.

Sthle, L., & Wold, S. (1989). Analysis of variance (ANOVA). *Chemometrics and intelligent laboratory systems*, *6*(4), 259-272.

Veloso, L., McHugh, J., von Weltin, E., Lopez, S., Obeid, I., & Picone, J. (2017, December). Big data resources for EEGs: Enabling deep learning research. In *2017 IEEE signal processing in medicine and biology symposium (SPMB)* (pp. 1-3). IEEE.
20